# Magnetic and transport properties of Mn$_2$CoAl oriented films

Michelle E. Jamer, Badih A. Assaf, Trithep Devakul and Don Heiman
Department of Physics, Northeastern University, Boston, MA 02115

The structure, magnetic and transport properties of thin films of the Heusler ferrimagnet Mn$_2$CoAl have been investigated for properties related to spin gapless semiconductors. Oriented films were grown by molecular beam epitaxy on GaAs substrates and the structure was found to transform from tetragonal to cubic for increasing annealing temperature. The anomalous Hall resistivity is found to be proportional to the square of the longitudinal resistivity and magnetization expected for a topological Berry curvature origin. A delicate balance of the spin-polarized carrier type when coupled with voltage gate-tuning could significantly impact advanced electronic devices.

Spin gapless semiconductors[1,2,3] (SGS) uniquely merge the properties of half-metallic magnets and zero-gap semiconductors. This remarkable combination of properties in SGS could have advantages for spin-transport electronic/magnetic devices and quantum information processing. New functionalities are poised to take advantage of several novel and valuable properties, even at room temperature. Some of the unique properties are: (1) half-metallic high spin polarization (~100 %); (2) carrier flexibility allowing for conduction of spin-polarized *holes* as well as spin-polarized electrons; (3) an ability to switch between spin-polarized *n*-type and spin-polarized *p*-type functionality by applying a gate voltage; (4) voltage-tunable spin polarization by varying the Fermi level; (5) some SGS compounds are predicted[3] to be half-metallic *antiferromagnets*. Several SGS materials have grown in bulk form,[2,4,5] but devices will require synthesis of thin films. The present study examines the growth of Mn$_2$CoAl in thin film form on GaAs substrates.

The density of states (DOS) as a function of energy for the two spin states of SGS is illustrated in Fig. 1(a). For one spin state (down spin), the Fermi level lies in an energy gap similar to half-metals, thus allowing for a high degree of spin polarization. In addition to that gap, the Fermi level for the other spin state (up spin) lies at a point where the DOS advantageously approaches zero for both conduction and valance bands.

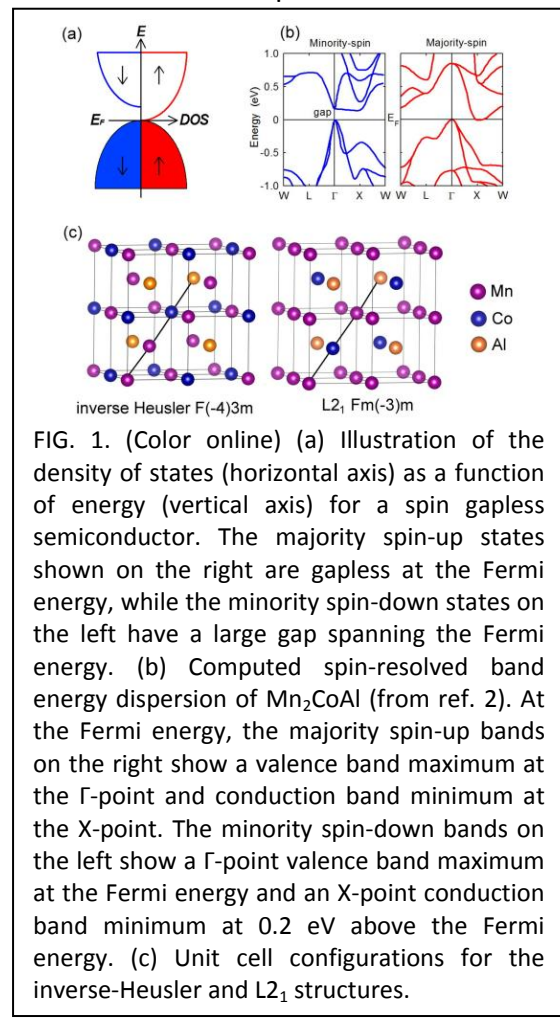

FIG. 1. (Color online) (a) Illustration of the density of states (horizontal axis) as a function of energy (vertical axis) for a spin gapless semiconductor. The majority spin-up states shown on the right are gapless at the Fermi energy, while the minority spin-down states on the left have a large gap spanning the Fermi energy. (b) Computed spin-resolved band energy dispersion of Mn$_2$CoAl (from ref. 2). At the Fermi energy, the majority spin-up bands on the right show a valence band maximum at the Γ-point and conduction band minimum at the X-point. The minority spin-down bands on the left show a Γ-point valence band maximum at the Fermi energy and an X-point conduction band minimum at 0.2 eV above the Fermi energy. (c) Unit cell configurations for the inverse-Heusler and L2$_1$ structures.

Thus far, more than half a dozen Heusler materials have been predicted to have SGS properties. These include $Mn_2CoAl$, which has an indirect band structure for the majority spin-up states, where the valence band touches the Fermi level at the Γ-point and the conduction band touches at the X-point, shown in Fig. 1(b). The magnetism is ferrimagnetic with total moment $2\mu_B$, where the Mn atoms on the two different sublattices are antialigned, but have different magnetic moments. A similar compound, $Ti_2CoSi$, is predicted to be ferromagnetic with a total moment of $3\mu_B$. Other SGS compounds, such as $Ti_2MnAl$ and $V_3Al$, are predicted to be antiferromagnetic.[3] The magnetic transition temperature of SGS materials is expected to be high, in the range 400 to 800 K, making them useful for room temperature device applications. Only a few materials have been synthesized with the aim of investigating the SGS properties. All of these are in bulk form and include $Mn_2CoAl$,[2,6] $CoFeMnSi$,[5] and $(Fe,V)_3Al$,[7] mostly grown by arc melting.

$Mn_2CoAl$ thin films were grown epitaxially on a GaAs substrate by MBE, which resulted in a tetragonal distortion. Annealing studies showed that the films lose their epitaxial registration and approach an oriented cubic structure after annealing, while increasing their crystallinity, chemical ordering and magnetic properties. The magnetic and magnetotransport properties were investigated for temperatures 2-400 K and fields to 14 T. The temperature-dependent resistivity shows metallic-like behavior below 200 K, but turns over into a semiconducting-like negative slope above 200 K. The magnetoresistance is negative and the Hall resistivity $\rho_{xy}$ is found to scale with the resistivity $\rho_{xx}^2$ and magnetization for all temperatures and magnetic fields, expected for a topological origin for the anomalous Hall effect (AHE) that is related to the Berry-phase curvature.[8,9]

Thin films of $Mn_2CoAl$ were grown by solid-source MBE on GaAs (001) substrates. Before growth, the GaAs was desorbed of the surface oxide by heating to 620 °C in an As flux having a pressure of $\sim 10^{-5}$ torr. Reflection high energy electron diffraction (RHEED) was used to monitor the desorption process and structure of the surface atoms during growth. The samples were capped with 2nm of Al to form an $AlO_x$ protective layer. RHEED patterns of the desorbed substrate and film during growth are shown in the inset of Fig. 2(b). The pattern of the $Mn_2CoAl$ film has twice the periodicity of the substrate, as the Heusler structure has 4 sublattices instead of the 2 sublattices of GaAs. From the spacing of the RHEED streaks the surface atoms appear to have the same lattice constant as the substrate to within 5 %, indicating epitaxial registration with the underlying substrate. Samples were grown at 100 to 200 °C then annealed for 30 min in vacuum at temperature from 200 to 400 °C.

Composition of the $Mn_2CoAl$ films was determined from energy dispersive spectroscopy (EDS). Among the several dozen samples grown on Si and GaAs, a number of samples were chosen that were within ±2% of the ideal composition. The structure and surface morphology was characterized by XRD measurements, atomic force microscopy (AFM), and scanning electron microscopy (SEM). Magnetic properties were investigated by superconducting quantum interference device (SQUID) magnetometry (Quantum Design XL-5). The magnetotransport was measured on Hall bars patterned with submillimeter dimensions. Electrical measurements were made in the magnetometer using a probe that was modified for conductivity measurements[10] in fields to 5 T and temperatures 2-400 K, and in a refrigerated cryostat placed in the room temperature bore of a cryogen-free 14 T magnet (Cryogenics Ltd.) at temperatures 6-300 K.

$X_2YZ$ full Heusler compounds contain 4 *fcc* sublattices that form in several crystal types depending on the sublattice ordering. Fully-ordered $Mn_2CoAl$ has a prototype $Hg_2TiCu$ XA (inverse Heusler) structure with a F(-4)3m space group identified by a 4-atom basis configuration of Mn-Mn-Co-Al along the diagonal, shown in the left diagram of Fig. 1(c).[11,12] This lattice is similar to the $L2_1$ lattice, characteristic of $Co_2MnZ$, which has a space group of Fm(-3)m and 4-atom basis configuration of Co-Mn-Co-Al along the diagonal, shown in the right diagram. Depending on the growth conditions the sublattices can be mixed, resulting in a B2 *bcc* structure for partial intermixing or an A2 *bcc* structure for complete intermixing.[13]

The crystal structure of the films was investigated by XRD as a function of the annealing conditions. XRD scans were made in a θ-2θ configuration using Cu Kα radiation at 1.5418 Å. Figure 2(a) shows the $Mn_2CoAl$ (002) and (004) Bragg diffraction peaks appearing on the low angle sides of the GaAs (002) and (004) substrate peaks, respectively. The (004) peak is shown in more detail in Fig. 2(b), where it shifts to higher angle for increasing annealing temperature. Figure 2(c) shows that the *c*-axis lattice constant decreases from 5.97 to 5.82 Å as the annealing temperature increases from $T_a$=200 to 325 °C. Applying the Scherrer formula to the width of the (004) peak results in a lower limit of D=13 nm (neglecting strain) for the coherent dimension in the perpendicular direction. For this range of temperatures no other significant XRD peaks are observed, demonstrating that the *c*-axis remains aligned perpendicular to the film plane. At the highest annealing temperature of $T_a$=400 °C the (004) peak is pushed into the tail of the substrate peak, as seen by the shoulder in Fig. 2(b), where it cannot be resolved. For the 400 °C annealing condition, the Mn (001), (002), (003) and (004) peaks become evident. The Mn (004) peak is seen on the low angle side of the $Mn_2CoAl$ (004) peak.

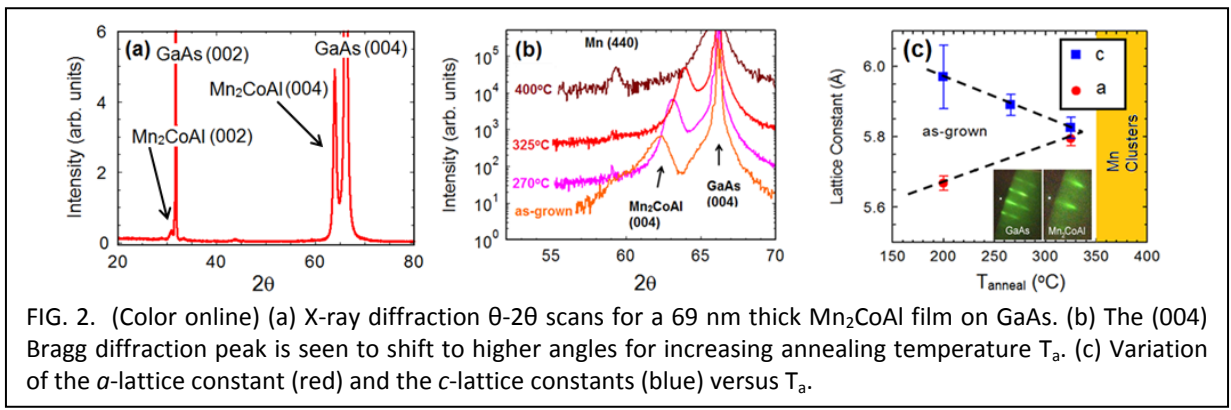

FIG. 2. (Color online) (a) X-ray diffraction θ-2θ scans for a 69 nm thick $Mn_2CoAl$ film on GaAs. (b) The (004) Bragg diffraction peak is seen to shift to higher angles for increasing annealing temperature $T_a$. (c) Variation of the *a*-lattice constant (red) and the *c*-lattice constants (blue) versus $T_a$.

In-plane diffraction was performed in order to determine the *a*-axis lattice constant as a function of annealing. The as-grown sample has an in-plane lattice parameter close to that of GaAs. $Mn_2CoAl$ (511) and the GaAs (511) peaks were observed. The as-grown film is thus epitaxial with the substrate and exhibits a tetragonal structure with *c*=5.97±0.09 Å, and an in-plane lattice constant $a=5.66\pm0.02$ Å that is close to the 5.653 Å GaAs lattice constant. For increasing annealing temperatures, the lattice transforms from a tetragonal structure with $c/a=1.05$ to a nearly cubic structure. At $T_a=325$ °C, the $a$- and $c$-axis lattice constants were found to be nearly identical within error, where $a=c=5.82$ Å. For this annealing temperature a small splitting between the (224) and (422) peaks is resolved, indicating a slight residual distortion from a cubic structure. The lattice constant of the annealed films is similar to those measured for bulk $Mn_2CoAl$, which range from 5.798[2] to 5.8546 Å.

The relative intensities of the (002) and (004) Bragg peaks were measured and compared to calculations in order to determine ordering of the atoms on the 4 sublattices.[11,14] First, we note that the unit cells of the film remain aligned with the substrate for the various annealing conditions at least up to 325 °C, as the expected (220) and (422) major powder peaks are absent. The ratio of intensities can be translated into a mixing order parameter given by $S=[I(002)/I(004)]^{1/2}/[I'(002)/I'(004)]^{1/2}$, where $I$ and $I'$ are the experimental and theoretical integrated intensities, respectively. Theoretical intensities were calculated using atomic scattering factors[15] by evaluating the square of the structure factors for each lattice type[16]. For $T_a=325$ °C annealing, the ratio $I(004)/I(002)=17.6$ was observed. Using $I'(004)/I'(002)=3.3$ as the theoretical value for perfect ordering leads to an experimental ordering of $S=0.43$. This simple analysis cannot determine the type of intermixing for the ternary compound[17]; however, it is more likely that the adjacent Mn and Co atoms, or all of the sublattices intermix in the $F(\bar{4})3m$ inverse Heusler structure.

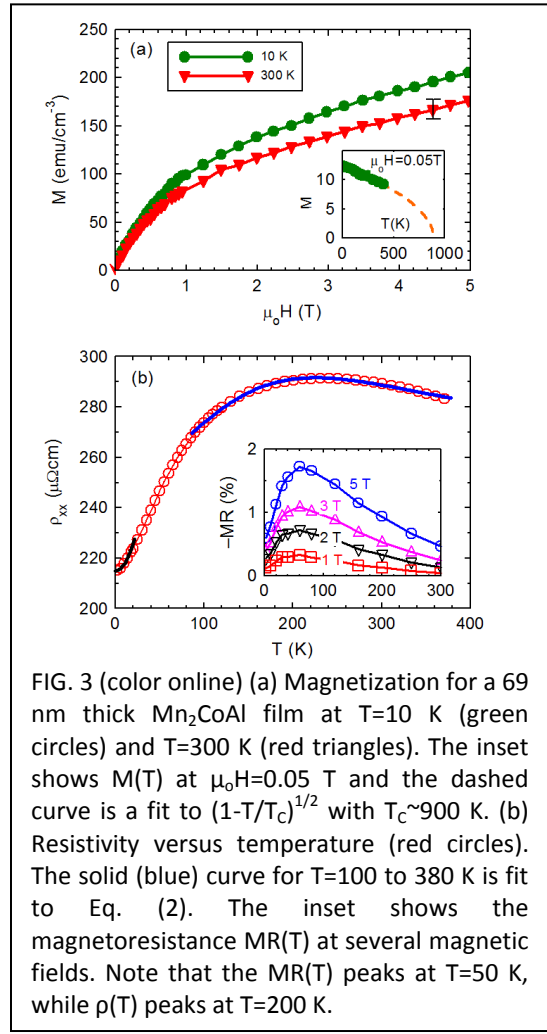

FIG. 3 (color online) (a) Magnetization for a 69 nm thick $Mn_2CoAl$ film at T=10 K (green circles) and T=300 K (red triangles). The inset shows M(T) at $\mu_oH=0.05$ T and the dashed curve is a fit to $(1-T/T_C)^{1/2}$ with $T_C\sim900$ K. (b) Resistivity versus temperature (red circles). The solid (blue) curve for T=100 to 380 K is fit to Eq. (2). The inset shows the magnetoresistance MR(T) at several magnetic fields. Note that the MR(T) peaks at T=50 K, while $\rho(T)$ peaks at T=200 K.

The magnetic properties of a $Mn_2CoAl$ film are shown in Fig. 3(a). The magnetization as a function field, M(H), is plotted up to $\mu_oH=5$ T. It has soft magnetic properties with a coercive field of $<10^{-2}$ T. The moment rises quickly at low fields but increases more slowly at high fields. The slow increase of M(H) is assigned to some antiferromagnetic exchange arising from the atomic disorder that was apparent from the XRD results. The predicted antiferromagnetic alignment of the Mn-Mn and Mn-Co neighbors in the ordered state is crucial and exchanging positions of the atoms can lead to strong frustration of some moments. The large internal antiferromagnetic exchange fields require very high magnetic fields to reach saturation. This M(H) behavior is similar to the slow spin alignment of Mn ions in the random antiferromagnet (Cd,Mn)Te,[18] arising from clusters of antiferromagnetically coupled Mn atoms. At 5 T a moment of $1\mu_B$ per formula unit is reached for our films, smaller than the $2\mu_B$ expected for a fully-ordered material and observed in arc melted bulk samples.[2,6] The Curie temperature, $T_C$, for the film is clearly well above 600 K, as the inset shows only a small decrease in M(T) up to 400 K, in agreement with the measured $T_C$=720 K in bulk-grown Mn$_2$CoAl.[2]

The electrical properties of Mn$_2$CoAl films were measured in the temperature range 2-375 K and in magnetic fields to 5 T. Figure 3(b) shows the temperature dependence of the zero-field resistivity, $\rho_{xx}$(T). The low-temperature resistivity is $\rho_{xx}$=220 µΩcm, falling between that of other bulk SGS compounds Fe$_2$CoSi (30 µΩcm)[5] and Mn$_2$CoAl (45 mΩcm)[2]. Those materials show a linear $\rho_{xx}$(T) dependence at temperatures above T=50 K. Here, a similar metallic-like behavior is seen at lower temperatures, but a maximum is observed at T~200 K, followed by a negative slope above room temperature. The temperature dependence of ρ(T) is contained in

$$\rho_{xx}(T) = \frac{1}{n(T)e\mu(T)}, \qquad \text{Eq. (1)}$$

where the carrier density (n) and mobility (µ) can vary with temperature. A positive slope in ρ(T) is associated with metallic-like behavior arising from constant n(T) and decreasing µ(T) due to increasing carrier-phonon scattering for increasing temperature. In contrast, a negative slope in ρ(T) can arise from several mechanisms. Impurities in semiconductors experience thermal activation that release mobile carriers leading to an increase in n(T). On the other hand, spin-disorder scattering in magnetic compounds can cause $\rho_{xx}$(T) to pass through a maximum at $T_C$. For example, (Fe,V)$_3$Al compounds exhibit a peak in ρ(T), resulting in negative resistivity slope at higher temperatures.[19] However, in the present Mn$_2$CoAl film, the resistivity maximum is at T~200 K, much lower than our $T_C$ that is well above 600 K. In addition, although spin-disorder scattering leads to our observed negative magnetoresistance, MR(H)=[$\rho_{xx}$(H)-$\rho_{xx}$(0)]/$\rho_{xx}$(0), it is unlikely that it gives rise to the ρ(T) maximum at 200 K, as the MR reaches a maximum at a much smaller temperature ~50 K, as shown in the inset of Fig. 3(b). The negative MR is a consequence of the reduction of random spin-flip scattering processes as the field aligns the magnetic moment. Bulk SGS materials have shown negative MR that transforms to positive MR at lower temperatures.[2,5] We also note that at the lowest temperatures ρ(T) deviates from the carrier-phonon scattering and exhibits a power law ~$T^\beta$, β=2.0±0.02, shown by the red curve in Fig. 3b, that is characteristic of electron-hole scattering in a degenerate semimetal.[20]

The behavior of ρ(T) in the region of the peak, from T=100 to 400 K, can be described by assuming a two-component model. In this model we consider the addition of conductivities, where the total conductivity is given by

$$-1/\rho_{xx} = \sigma_{xx}(T) = n_m e \mu_{ph}(T) + n_a(T) e \mu_a(T). \qquad \text{Eq. (2)}$$

The two contributions act as independent parallel conduction channels. We assign the first term to the metallic-like contribution, where $n_m$ is independent of temperature and the mobility $\mu_{ph}$(T)=1/(cT+d) is due to carrier-phonon scattering. The second term is assigned to the semiconducting-like contribution, where $n_a(T) = n_o e^{-\Delta/T}$ provides thermally activated carriers with activation temperature Δ. An activated contribution would result from the freeze-out of carriers from an impurity state characterized by an energy barrier Δ or a gapped semiconductor state. At low temperatures where the ρ(T) slope is positive the second term diminishes and ρ(T) is governed by the phonon-dominated mobility. At high temperatures where the slope is negative the second term increases and ρ(T) is governed by the thermally-released carriers. In order to fit the data we simplify the problem by assuming that the mobility is the same for both types of carriers, $\mu_a(T)=\mu_{ph}(T)$. Results of the fit to Eq. (2) are shown by the solid (blue) curve in Fig. 3(b), where the activation temperature is Δ=720 K (62 meV). In view of these results, the behavior of $\rho_{xx}(T)$ could arise from a two-carrier electron-hole regime where the conduction and valence bands overlap and a thermally active state lies 60 meV below the Fermi energy.

Results from Hall effect measurements on a $Mn_2CoAl$ film are shown in Fig. 4. The Hall resistivity, $\rho_{xy}(H,T)$, is shown by solid triangles for fields up to $\mu_o H$=5 T at T=300 K. The anomalous Hall effect (AHE) is analyzed by the following method, using

$$\rho_{xy}(H) = R_0 H + R_1 M(H). \qquad \text{Eq. (3)}$$

Here, $R_0=1/(ne)$ is the ordinary Hall effect (OHE) coefficient for a carrier density $n$, and $R_1=a\rho_{xx}+b\rho_{xx}^2$ is the anomalous Hall coefficient, where $a$ is the coefficient for skew scattering,[21,22] and $b$ is the coefficient for combined side-jump scattering and intrinsic topological Berry phase contributions.[8] Equation (3) has explicit H-dependence since the magnetization does not reach saturation. It was found that the $a$-coefficient for skew scattering was negligible and was set to $a$=0. Equation (3) was fit to the measured $\rho_{xy}(H)$ in order to determine the fitting parameters $R_0$ and $b$, at the four temperatures T=10, 100, 200, 300 K. Figure 4 shows the measured $\rho_{xy}(H)$ for T=300 K, along with the fitted results for $R_0 H$ and $R_1 M(H)$. The OHE shows a small negative-going linear behavior and the AHE behavior is similar to M(H). The sum $R_0 B+R_1 M(H)$ is displayed by the solid curve and shows excellent correspondence to $\rho_{xy}(H)$. Fits to the other three temperatures also show excellent agreement.

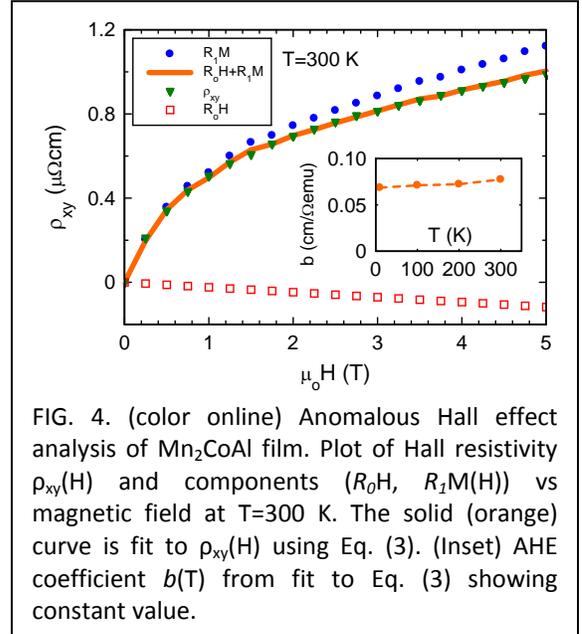

FIG. 4. (color online) Anomalous Hall effect analysis of $Mn_2CoAl$ film. Plot of Hall resistivity $\rho_{xy}(H)$ and components ($R_0 H$, $R_1 M(H)$) vs magnetic field at T=300 K. The solid (orange) curve is fit to $\rho_{xy}(H)$ using Eq. (3). (Inset) AHE coefficient $b(T)$ from fit to Eq. (3) showing constant value.

The inset of Fig. 4 shows that $b(T)$ is nearly independent of T, and thus demonstrates that the AHE *conductivity* is proportional to M(H,T), where $\sigma_{xy}^{AHE}(H,T)=\rho_{xy}^{AHE}(H,T)/\rho_{xx}^2=bM(H,T)$. We also note that since MR(H) is rather small, <3%, compared to changes in $\rho_{xx}(T)$, the same values for $b$ are obtained by using either $\rho_{xx}(H,T)$ or $\rho_{xx}(H=0,T)$ in the analysis. The observed scaling law $\rho_{xy}=bM/\rho_{xx}^2$ indicates that the AHE has an intrinsic origin or is related to the side-jump scattering. However, the present value of $\rho_{xx}$~200-300 μΩcm favors a topological origin of the AHE that can be computed from the Berry-phase curvature.[8,9] Our value for the Hall conductivity, $\sigma_{xy}$=12 S/cm measured at 5T, lies between the value measured for bulk Mn$_2$CoAl, 22 S/cm, and that computed from the Berry phase curvature where 3 S/cm.[2] All of these values are much smaller than that found in typical ferromagnets, thus favoring an intrinsic origin for the AHE.

The electron concentration derived from $R_0$(T) is large, $n=4\pm2\times10^{22}$ cm$^{-3}$. This value is much larger than observed in bulk Mn$_2$CoAl,[2] where $n=2\times10^{17}$ cm$^{-3}$, presumably due to differences in atomic ordering. It is important to note that in our analysis, even a large error of 1% in subtracting the substrate diamagnetism would change $n$ by ~25%. There is the possibility of two types of carriers, holes and electrons. A two-carrier system is realistic considering that both conduction and valence bands are expected to be close to the Fermi energy,[2,3] as shown in Fig. 1(b). The bands may overlap in energy when crystal defects are introduced. Also, the bands near the Fermi energy are at different points in the Brillouin zone, which could naturally lead to separate electron and hole pockets in the Fermi surface.

In conclusion, extending the synthesis of SGS from bulk to thin films is crucial for the development of these new materials for electronic devices. It was found that films of Mn$_2$CoAl can be grown by MBE on GaAs substrates yielding high crystalline alignment. The as-grown films were found to have a tetragonal distortion due to the epitaxy, but post-growth annealing transforms the films into a cubic structure while retaining crystalline alignment. XRD and magnetization results indicated disorder in the atomic sublattices. Magnetotransport measurements on annealed films revealed a semiconducting-like $\rho_{xx}$ resistivity, Hall resistivity proportional to $\rho_{xx}^2$ and M(H,T), and small Hall conductivity representative of an intrinsic topological AHE. Although the present films were not fully chemically ordered, future work on SGS films will need to address the issues of stoichiometry and chemical ordering of the sublattices, which is more challenging than in bulk samples.

**Acknowledgements** - This work supported by the National Science Foundation grant DMR-0907007. We thank T. Hussey for assistance with experimental apparatus and F. Jimenez-Villacorta for assistance with XRD experiments.